# EVALUATING RACE AND SEX DIVERSITY IN THE WORLD'S LARGEST COMPANIES USING DEEP NEURAL NETWORKS

Konstantin Chekanov[1], Polina Mamoshina[2,3], Roman V. Yampolskiy[4], Radu Timofte[5], Morten Scheibye-Knudsen[6*], Alex Zhavoronkov[2,7,8*]

[1] Youth Laboratories, Ltd, Diversity AI Group, Skolkovo Innovation Center, Nobel Street 5, 143026, Moscow, Russia

[2] Insilico Medicine, Emerging Technology Centers, JHU, 1101 33rd Street, Baltimore, MD, 21218, USA

[3] Computer Science Department, University of Oxford, Oxford, United Kingdom

[4] Computer Engineering and Computer Science, Duthie Center for Engineering, University of Louisville, Louisville, KY 40292, USA

[5] Computer Vision Lab, Department of Information Technology and Electrical Engineering, ETH Zürich, Switzerland

[6] Center for Healthy Aging, Department of Cellular and Molecular Medicine, University of Copenhagen, Denmark

[7] The Biogerontology Research Foundation, 2354 Chynoweth House, Trevissome Park, Truro, TR4 8UN, UK.

[8] Moscow Institute of Physics and Technology, Institutskiy per., 9, Dolgoprudny, 141701, Russia

\* Corresponding Authors:
Morten Scheibye-Knudsen: mscheibye@sund.ku.dk
Alex Zhavoronkov: alex@biogerontology.org

**Abstract**

Diversity is one of the fundamental properties for survival of species, populations and organizations. Recent advances in deep learning allow for the rapid and automatic assessment of organizational diversity and possible discrimination by race, sex, age and other parameters. Automating the process of assessing the organizational diversity using the deep neural networks and eliminating the human factor may provide a set of real-time unbiased reports to all stakeholders. In this pilot study we applied the deep learned predictors of race and sex to the executive management and board member profiles of the 500 largest companies from the 2016 Forbes Global 2000 list and compared the predicted ratios to the ratios within each company's country of origin and ranked them by the sex-, age- and race- diversity index (DI). While the study has many limitations and no claims are being made concerning the individual companies, it demonstrates a method for the rapid and impartial assessment of organizational diversity using deep neural networks.

**Abbreviations**

AI – artificial intelligence; b(%) and o(%) – percentage of African black and other races, respectively; CNN – convolutional neural network, DL – deep learning; DNN – deep neural networks; f(%) and m(%) – percentage of females and males, respectively; r – Pearson correlation coefficient; $r_{crit}$ – critical value of the Pearson correlation coefficient; TOP500 Forbes – list of the 500 largest companies in the Forbes 2016 Global 2000; x(%) and y(%) – percentage of people below 60 years old and people above 60 years old, respectively.

**Introduction**

Deep learning (DL), a branch of artificial intelligence (AI) research, is rapidly gaining popularity and credibility (LeCun et al., 2015). Systems utilizing deep learning technology have outperformed humans in tasks ranging from image (Karpathy and Fei-Fei, 2015) and voice recognition (Graves et al., 2013) to classifying skin cancer (Esteva et al., 2017). Recent advances in computational methods have made it possible to create sufficiently complex genetic algorithms (LeCun et al., 2015). Deep neural networks (DNNs) are widely used for facial recognition and are capable of accurately predicting the sex and age of photographed subjects (Zhao and Chellappa, 2002; Khalajzadeh et al., 2014; Sun et al., 2015). These results have spurred the development of DNN systems capable of predicting a patient's chronological age and medical state by analyzing biomarkers in patient blood samples (Putin et al., 2016). Moreover, DL-based systems enable machines to engage in creative activity – traditionally held as a uniquely human capability – such as poetry, painting and musical composition (Olivera, 2009, Boden, 2014). Similar systems have even been used to identify and describe attractive human faces and their attributes (Wong et al., 2008).

Despite the spectacular progress and achievements made in the field of AI research, the discipline is far from faultless. It has been reported that AI systems occasionally develop unfavorable human characteristics such as discriminatory behavior. For example, DeepID2+, a high-performance deep convolutional neural network (CNN) developed by Sun et al. (2015), is capable of distinguishing age-, sex- and race-related characteristics via its face recognition algorithm. Consequently, a diverse dataset is essential for the neural network's functionality and ability to discern identity. Upon closer inspection, the factors responsible for the emergence of prejudicial or discriminatory attributes, which are not intrinsic to the network, become obvious. In 2014, researchers developed an algorithm which calculates the probability that an individual will commit a repeated criminal offense (Angwin et al., 2016). Subsequent investigation revealed that the algorithm was significantly more likely to assign a greater probability of recidivism to black defendants than white defendants. Similarly, attempts to teach AI to recognize linguistic constructions have resulted in the emergence of AI-generated race- or sex-related stereotypes similar to those observed in humans (Caliskan et al., 2017).

Presently, the fight against different forms of discrimination is led by governmental action, particularly in North American and European countries (Pager, 2008; Becker, 2010; Caldera, 2013). The desire to discourage discrimination is enshrined in both national and international law; many nations introduced the anti-discrimination legislation in addition to the numerous international antidiscrimination treaties. Beyond the governmental response, there is evidence to suggest that discrimination perpetrated by an employer negatively affects the profit and reputation of the company (Pager, 2008; Becker, 2010). Despite this negative reinforcement, it is important to recognize the continued existence of discrimination in the economic domain (Pager, 2008; Robinson and Dechant, 1997; Blanchflower et al., 2003). In addition to economic disadvantages, individuals and groups that experience discrimination may be at greater risk for developing medical conditions or diseases (Krieger, 1990). Counterintuitively, it is possible that the introduction of antidiscrimination laws has unveiled the latent forms of discrimination (Pager, 2008). Three abundant forms of such discrimination are: i) by sex (sexism), ii) by age (ageism), or iii) by race (racism).

Is it possible for AI to present prejudicial biases? If so, it is important to determine if the source of the bias is an integral algorithmic component or is instead an artifact of human biases present in the data from which the AI learns. In order to elucidate this problem, we created a dataset that one might use to develop an algorithm that predicts human success (in this case, success in business) by analyzing images of the subject's face. Utilizing deep convolutional network algorithms, we analyzed the dataset in order to detect the presence of people belonging to groups that often experience discrimination. For this analysis, our dataset was populated by images of successful business executives, a dataset where we speculate that sexism, ageism and racism could occur. Specifically, we compiled publicly available photographs of the board of directors for the 500 largest companies in the Forbes 2016 Global 2000, which we denote as TOP500 Forbes. To facilitate analysis of the data, we calculated numerical indices corresponding to the three most prevalent categories of discrimination (sex, age and race) and compared the values for different countries of the world.

## Methods

### The general scheme of the analysis

An overview of the analysis is shown in Fig. 1. It included companies selection, identification of the country of origin (and state in the case of United States), a linking to top management profiles, identification of the pictures of top managers, running the pictures through DL predictors of chronological age, race and sex, comparing the resulting values with the values in country of origin and the diversity indices calculation.

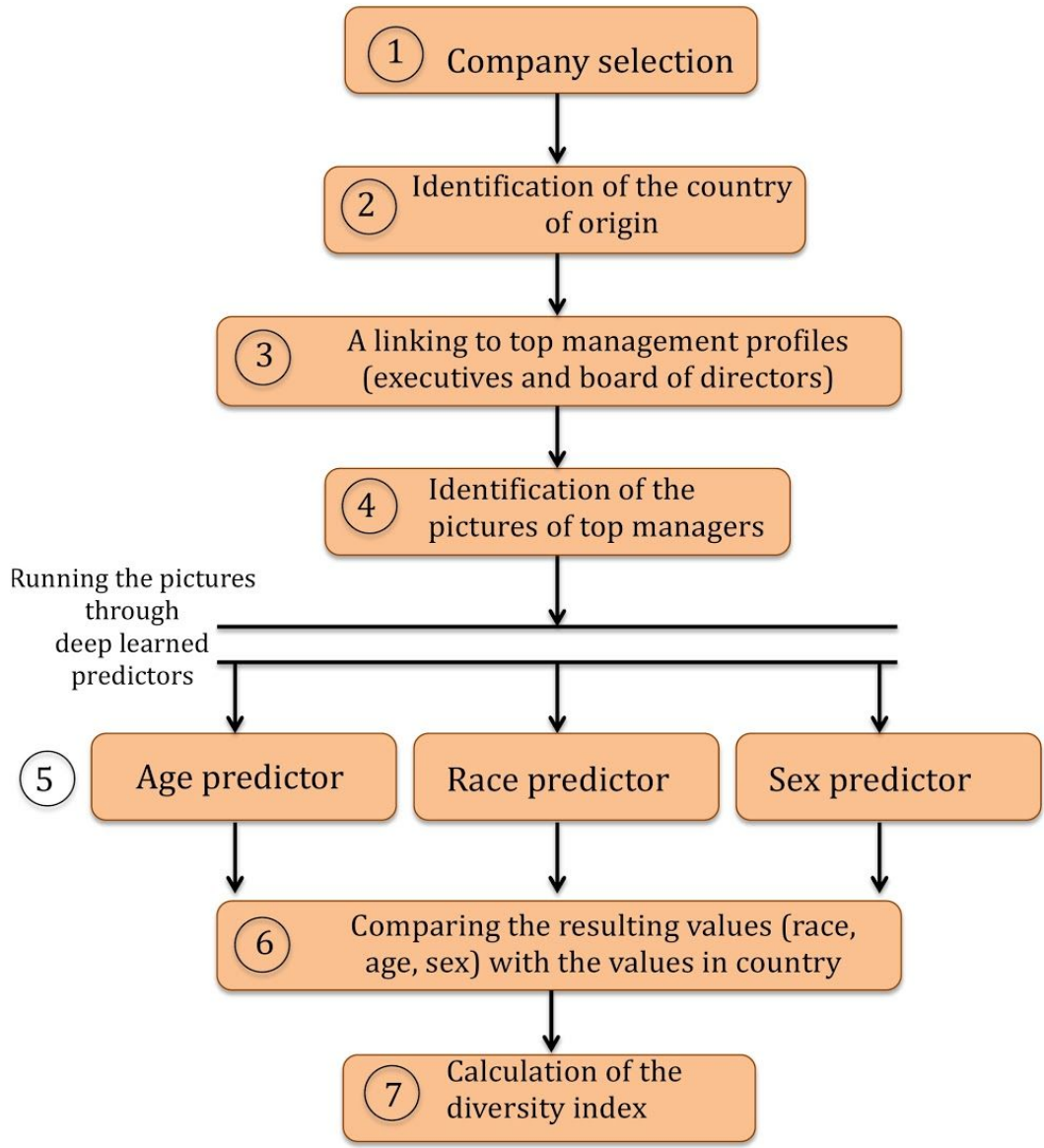

Figure 1: The general scheme of the analysis: (1) companies selection → (2) identification of the country of origin (and state in the case of United States) → (3) a linking to top management profiles → (4) identification of the pictures of top managers → (5) running the pictures through DL predictors of chronological age, race and sex (in parallel) → (6) comparing the resulting values with the values in country of origin → (7) the diversity indices calculation.

**The dataset**

Photographs, chosen based on information listed on the Forbes website of corporate executives from the 500 largest companies (Forbes 2016 Global 2000 ranking), were collected from the companies' official websites and saved in JPG or PNG format. The dataset was compiled on March 20th, 2017. In this work, we use the term "corporate executive" to refer to positions including board of directors, executive management team, executive directors, non-executive directors, executive officers, president of the company, supervisory board, senior management, audit board supervisory, audit committee, management committee, board of supervisors, key executive team, and senior executive team. If the photo found on the official website was of insufficient quality, a high-resolution publicly available image would be used as a substitute. Each photo was labeled by the name of the company, by the sex and race (Fig. 2). All companies were categorized by location as European, Asian, American or other companies (including African and Australian).

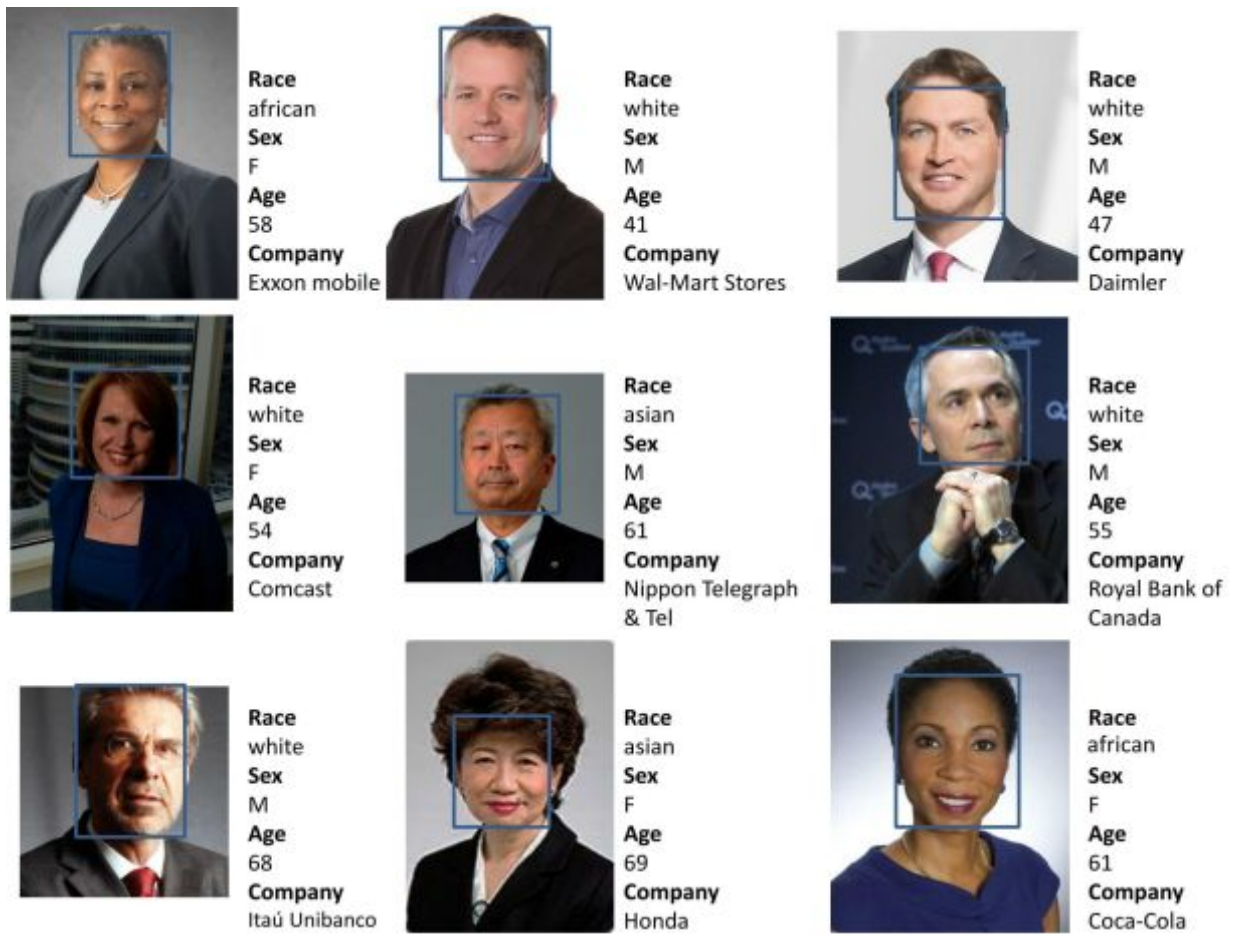

Figure 2: Publicly available sample photographs of the corporate executives of the 500 largest companies in the Forbes 2016 Global 2000 ranking. Each photo was labeled by the name of the company, sex, race and age. Preliminary face detection was carried out for the age, race and sex prediction.

For many companies including Schlumberger Limited lacking the pictures linked to the executive profiles, we performed a basic search for profile pictures using the Google Image Search (images.google.com) and LinkedIn (www.LinkedIn.com) images. The probability is high that some of the images may have been mismatched.

**Sex, age and race prediction**

The sex, race and age were predicted using deep CNNs. The prediction by trained CNNs was carried out using the Caffe framework (Jia et al., 2014).

For age prediction, the CNN with VGG-16 architecture from Rothe et al. (2016) was used. Reported CNN was pre-trained on ImageNet for image classification and then trained on a IMDB-WIKI dataset of facial images with age and gender labels (Rothe et al., 2016). There were a total of 101 neurons for age prediction in the output layer with a softmax function (corresponding to the age from 0 to 100 years). The lowest reported mean absolute error of age estimation was of 3.221 years (Rothe et al., 2016).

For race prediction, we used CNN DeepID2+ (Sun et al., 2015) with 4 convolutional layers, followed by a feature layer fully-connected to third and fourth layers. The reported accuracy of face features recognition was 93.2 ± 0.2. The reported network was able to handle corrupted images and in our work demonstrated the best performance in terms of race recognition. In the present study three races were predicted: Asian, Black and White.

For sex prediction, CNN classifier from Levi and Hassner (2015) with three convolutional layers followed by a linear rectified, pooling layers, two fully-connected layers and softmax function was used. The average reported accuracy of sex determination was 86.8±1.4 ( Levi and Hassner, 2015).

**Results**

**Total dataset**

The final dataset consisted of 7207 face photographs of corporate executives from 484 companies listed in the Forbes 2016 Global 2000. Images of the corporate executives for 16 companies were not publicly available at the time of data collection. The 484 companies studied were located in 38 countries, as shown by the distribution in Table 1.

**Table 1. The distribution of the companies from TOP500 Forbes by country.**

| Country | Number of Companies | Country | Number of Companies |
|---|---|---|---|
| Europe (147 companies from 18 countries) | | | |
| UK | 29 | Sweden | 8 |
| Germany | 22 | Israel | 1 |
| France | 26 | Denmark | 3 |
| Switzerland | 15 | Ireland | 2 |
| Spain | 8 | Norway | 1 |
| Netherlands | 9 | Finland | 2 |
| Russia | 6 | Austria | 1 |
| Belgium | 2 | Portugal | 2 |
| Italy | 7 | Turkey | 3 |
| Asia (129 companies from 10 countries) | | | |
| Japan | 41 | Hong-Kong | 9 |
| Korea | 11 | Taiwan | 4 |
| Singapore | 5 | Thailand | 1 |
| Malaysia | 2 | China | 41 |
| India | 13 | Indonesia | 2 |
| America (192 companies from 4 countries) | | | |
| USA | 173 | Canada | 13 |
| Brazil | 4 | Mexico | 2 |
| Other (16 companies from 5 countries) | | | |
| Australia | 8 | Saudi Arabia | 3 |
| UAE | 1 | the Republic of South Africa | 3 |
| Qatar | 1 | | |

The median, $25^{th}$ quartile and $75^{th}$ quartile for predicted ages were 52, 42 and 52 years respectively. Females represented 21.2 % of the total number of corporate executives photographed. Of the photographed corporate executives, 79.7 % were White (Caucasian), 3.6 % were Black and 16.7 % were Asian.

**Sex**

The sampled companies with the highest predicted percentage of females among their corporate executives are presented in Table 2. It should be noted that, the percentage of female corporate executives was lower than the average percentage of able-bodied females in the country's population for almost all companies, except H&M from Sweden. The larger values in the dataset were found to correspond to companies located in Europe and the United States. This may be due to a higher average percentage of able-bodied females in the population as compared to other countries. Of the 484 sampled companies, at least 23 did not have any female corporate executives whose photographs were available (Table 3). A substantial majority of the companies without female corporate executives were located in Asia, with a large number of them in China and Japan. The average accuracy of sex determination was 75.5±2.5.

The values of r between the predicted percentage of women on an executive position for the companies with maximum female inclusion and a total percentage of working females in countries was 0.21, which is lower than its critical value ($r_{crit}$ of 0.47 for confidence interval (P) of 0.95 and number of samples (n) of 18).

**Table 2. The companies from TOP 500 Forbes with the highest predicted female representation. The percent of the able-bodied female population is also presented for each country (according to worldbank.org statistics for 2016).**

| Company | Country | Region | Predicted % of females in the company | Actual % of females in the company | Labor force, female (% of total labor force) |
|---|---|---|---|---|---|
| H&M | Sweden | Europe | 58 | 58 | 47 |
| Air Liquide | France | Europe | 54 | 42 | 48 |
| BNP Paribas | France | Europe | 47 | 47 | 48 |
| L'Oréal Group | France | Europe | 47 | 47 | 48 |
| Target | USA | America | 46 | 42 | 46 |
| Unilever | Netherlands | Europe | 46 | 39 | 46 |
| Synchrony Financial | USA | America | 44 | 33 | 46 |
| Poste Italiane | Italy | Europe | 44 | 31 | 42 |
| TJX Cos | USA | America | 44 | 36 | 46 |
| ING Group | Netherlands | Europe | 43 | 27 | 46 |
| TD Bank Group | Canada | America | 43 | 33 | 47 |
| Sanofi | France | Europe | 42 | 42 | 48 |
| eBay | USA | America | 42 | 25 | 46 |
| Citigroup | USA | America | 41 | 27 | 46 |
| Ericsson | Sweden | Europe | 41 | 41 | 47 |
| Royal Dutch Shell | Netherlands | Europe | 42 | 26 | 46 |
| Natixis | France | Europe | 40 | 43 | 48 |
| Macquarie Group | Australia | Other | 40 | 29 | 46 |

**Table 3. The companies from TOP 500 Forbes with zero predicted female corporate executives.**

| # | Company | Country | # | Company | Country | # | Company | Country |
|---|---|---|---|---|---|---|---|---|
| 1 | PetroChina | China | 9 | Tencent Holdings | China | 17 | Fujifilm Holdings | Japan |
| 2 | China Mobile | China | 10 | Nippon Steel & Sumitomo Metal | Japan | 18 | Metallurgical Corp of China | China |
| 3 | SoftBank | Japan | 11 | Nokia | Finland | 19 | EDP-Energias de Portugal | Portugal |
| 4 | Nissan Motor | Japan | 12 | Chubu Electric Power | Japan | 20 | KDDI | Japan |
| 5 | Hyundai Motor | Korea | 13 | Samsung Electronics | Korea | 21 | Denso | Japan |
| 6 | China State Construction Engineering | China | 14 | Saudi Telecom | Saudi Arabia | 22 | Hyundai Mobis | Korea |
| 7 | China Communications Construction | China | 15 | SK Hynix | Korea | 23 | Etisalat | UAE |
| 8 | Dalian Wanda Commercial Properties | China | 16 | Samsung Electronics | Korea | | | |

**Race**

For the purposes of this paper, we took the percentage of black corporate executives as the primary measure of racial discrimination. The majority of the companies with the highest percentage of black corporate executives was represented in the dataset of American companies (Table 4). It should be noted that even among the companies with the highest index, black executives comprised less than 20 % of the corporate executive personnel of most companies (Fig. 3D). This trend occurs independent of the racial demographics of the state in which the company is located. Furthermore, many companies located in the United States have zero black corporate executives – a trend that extends to other countries located in the Americas. This is clearly demonstrated by the total absence of black corporate executive personnel in each sampled company located in Brazil and Mexico, and the majority of the sampled companies located in

Canada (Table 4).

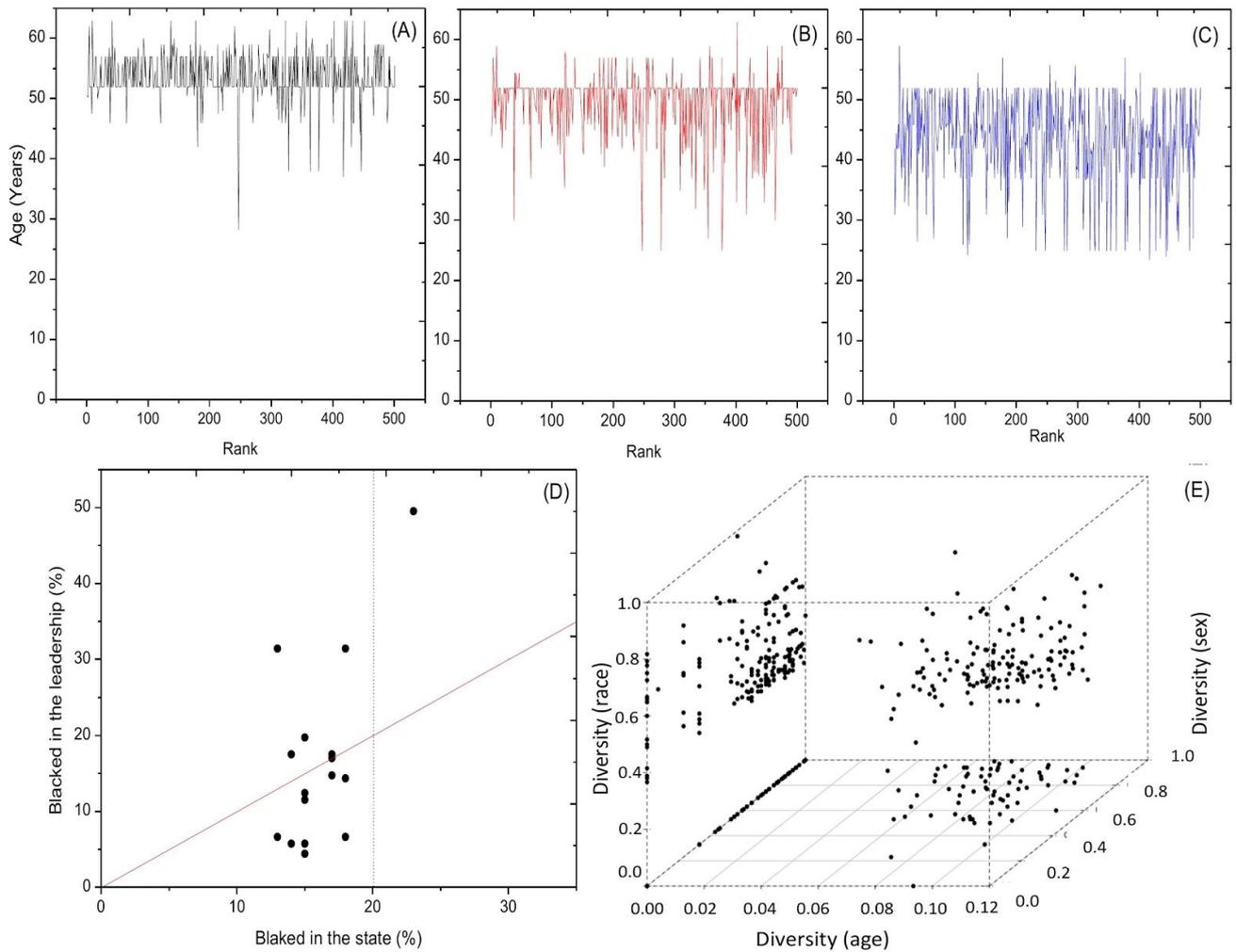

Figure 3: The 1st quartile (A), median (B) and 3rd quartile (C) of the predicted age of corporate executives of the 500 largest companies in the Forbes 2016 Global 2000 (TOP500 Forbes) (rank of the companies in the range of 1 - 500 was presented on the X-axis). The percentage of African American in the American companies from TOP500 Forbes with its highest value and the percentage of African American in the state of company's origin (where its headquarters is located) (D).

While the black population represents a low proportion of the total population in most European countries compared to the United States, African black executives were found to comprise a relatively substantial percentage of the total corporate executive personnel in some of the sampled European companies. Conversely, none of the corporate executive personnel in any sampled Asian company were predicted to be black. In general, Asian companies were characterized by a substantial majority of white and Asian corporate executives. For some companies located in China and Japan, this majority extended to totality. The sampled companies with headquarters in Saudi Arabia, UAE and Australia had zero black corporate

executives

**Table 4. The companies from TOP 500 Forbes with the largest predicted percentage of black corporate executives. The percentage of the able-bodied population identified as black is presented. For the United States, the percentage in each state is specified.**

| Company | Country | Predicted, % | Actual, % | Actual in the country |
|---|---|---|---|---|
| America | | | | |
| Fannie Mae | USA (District of Columbia) | 23 | 23 | 13 (46)[a] |
| Delta Air Lines | USA (Georgia) | 18 | 15 | 13 (31)[a] |
| Ally Financial | USA (Michigan) | 18 | 9 | 13 (14)[a] |
| Chevron | USA (California) | 18 | 14 | 13 (6)[a] |
| Verizon Communications | USA (New York) | 17 | 17 | 13 (14)[a] |
| Pfizer | USA (New York) | 17 | 17 | 13 (14)[a] |
| FedEx | USA (Tennessee) | 17 | 17 | 13 (17)[a] |
| Illinois Tool Works | USA (Illinois) | 17 | 9 | 13 (14.7)[a] |
| Archer Daniels Midland | USA (Illinois) | 17 | 18 | 13 (14.7)[a] |
| Target | USA (Minnesota) | 15 | 15 | 13 (6)[a] |
| Northrop Grumman | USA (Colorado) | 15 | 14 | 13 (4)[a] |
| ExxonMobil | USA (Texas) | 15 | 15 | 13 (12)[a] |
| Norfolk Southern | USA (Virginia) | 15 | 17 | 13 (19)[a] |
| PNC Financial Services | USA (Pennsylvania) | 15 | 15 | 13 (10)[a] |
| General Mills | USA (Minnesota) | 14 | 16 | 13 (6)[a] |
| American Express | USA (New York) | 14 | 14 | 13 (14)[a] |
| McKesson | USA (California) | 13 | 11 | 13 (6)[a] |
| Southern | USA (Georgia) | 13 | 13 | 13 (31)[a] |
| Europe | | | | |
| Eaton | Ireland | 17 | 17 | 1.4[b] |
| Old Mutual | UK | 13 | 14 | 3[b] |
| Accenture | Ireland | 10 | 9 | 1.4[b] |
| Unilever | Netherlands | 8 | 13 | - |
| Barclays | UK | 8 | 5 | 3[b] |
| Vivendi | France | 7 | 6 | 3-8[b] |
| Novartis | Switzerland | 7 | 7 | < 1[b] |
| National Grid | UK | 6 | 8 | 3[b] |
| Chubb | Switzerland | 5 | 5 | < 1[b] |
| Merck | Germany | 4 | 8 | < 1[b] |
| Danone | France | 4 | 5 | 3-8[b] |
| Nestle | Switzerland | 4 | 5 | < 1[b] |
| Vodafone | UK | 4 | 5 | 3[b] |
| Swiss Re | Switzerland | 2 | 2 | < 1[b] |
| Asia | | | | |
| NONE | | | | |
| Other | | | | |
| Sasol | the Republic of South Africa | 53 | 53 | 80[c] |
| Standard Bank Group | the Republic of South Africa | 35 | 35 | 80[c] |

[a]: according to http://www.kff.org, [b]: according to http://ec.europa.eu/, [c]: according to Henrard, 2000.

**Table 5 American companies from TOP 500 Forbes with zero predicted black corporate executives.**

| # | Company | Country | # | Company | Country |
|---|---|---|---|---|---|
| 1 | USA | Schlumberger | 22 | USA | Phillips 66 |
| 2 | USA | CRH | 23 | Mexico | América Móvil |
| 3 | USA | NextEra Energy | 24 | Brazil | Banco do Brasil |
| 4 | USA | Cisco Systems | 25 | USA | Qualcomm |
| 5 | Canada | TD Bank Group | 26 | USA | Philip Morris International |
| 6 | USA | The Priceline Group | 27 | USA | Abbott Laboratories |
| 7 | USA | MetLife | 28 | Canada | Sun Life Financial |
| 8 | USA | Walgreens Boots Alliance | 29 | Mexico | Femsa |
| 9 | USA | eBay | 30 | USA | Corning |
| 10 | USA | Berkshire Hathaway | 31 | USA | Hilton Worldwide Holdings |
| 11 | USA | Bank of New York Mellon | 32 | USA | EMC |
| 12 | Canada | Manulife Financial | 33 | USA | United Continental Holdings |
| 13 | USA | Biogen Idec | 34 | Canada | Power Corp of Canada |
| 14 | USA | CenturyLink | 35 | USA | Las Vegas Sands |
| 15 | USA | MasterCard | 36 | USA | Johnson Controls |
| 16 | USA | PayPal | 37 | USA | ConocoPhillips |
| 17 | Canada | Bank of Montreal | 38 | USA | Altria Group |
| 18 | Canada | Brookfield Asset Management | 39 | USA | Edison International |
| 19 | Canada | BCE | 40 | Brazil | Petrobras |
| 20 | USA | Paccar | 41 | Canada | Canadian Imperial Bank |
| 21 | Brazil | Banco Bradesco | 42 | Brazil | Itaú Unibanco Holding |
|  |  |  | 43 | USA | Intel |

**Age**

The median predicted age, as well as 1$^{st}$ and 3$^{rd}$ quartile ages, are presented in Fig. 3 (A, B, C). On average, the age predictions made by the algorithm fell below the subject's actual age, a tendency that was more prevalent among Asian subjects. For example, in some cases (data not shown) the algorithm predicted that the person was younger than 30 years old, but the youngest corporate executive among the sampled companies was found to be 32 years old. The actual chronological age for some executives of the sampled corporate executives was collected via the websites https://www.thomsonreuters.com and http://www.morningstar.com.

**Diversity index**

In order to represent the diversity found among corporate executives in the sampled companies, we introduced the diversity indices for age, sex and race. These indices of race diversity were defined as the modified reciprocal Simpson diversity index (Simpson, 1949), which is widely used in ecology in order to access the level of biological diversity:

$$\text{diversity index (race)} = 1/(M \sum_{i=1}^{M} (\frac{n_i}{N})^2) \times (1 - \sum_{i=1}^{M} \sum_{\substack{j \neq i}}^{M-1} \delta_{n_i,0} \delta_{n_j,0}),$$

where $n_i$ is a number of individuals identified as asian, black or white, respectively, $N$ is a total number of individuals in the company and $M$ is a number of races used for the analysis (in the present study $M$ equals 3).

$$\text{diversity index (sex)} = \frac{\sqrt{f(\%)m(\%)}}{50},$$

where $f(\%)$ and $m(\%)$ are the percentage of females and males, respectively, and

$$\text{diversity index (age)} = \frac{\sqrt{x(\%)y(\%)}}{50},$$

where $x(\%)$ and $y(\%)$ are the percentage of people < 60 years old and people ≥ 60 years old, respectively.

The values of these indices range between 0 and 1, where a higher value corresponds to a greater diversity. In this sense, the indices are inversely proportional to the disparity between input values, i.e. the index value approaches 1 as the parameter groups approach equality. The values of each of the three indices were calculated for every sampled company. The values can be seen in the distribution of the companies by diversity indices on Fig. 3E. Distribution of median diversity sex, race and age indexes by countries is presented on Fig. 4.

Based on the distribution of diversity index values, it is possible to identify the apparent biases of the sampled companies and categorize them accordingly. Companies which were found to have relatively high values for all three diversity indices can be labeled as "bias-free" companies, whereas companies which demonstrated a value of zero for any diversity index can be classified as "biased companies". Many of the companies not embodied by either of these antipodal categories fall into a group of companies which have high sex and age diversity index scores, but considerably lower racial diversity index scores – this is indicative of more pronounced racial bias as compared to biases related to age or sex (fig. 3E).

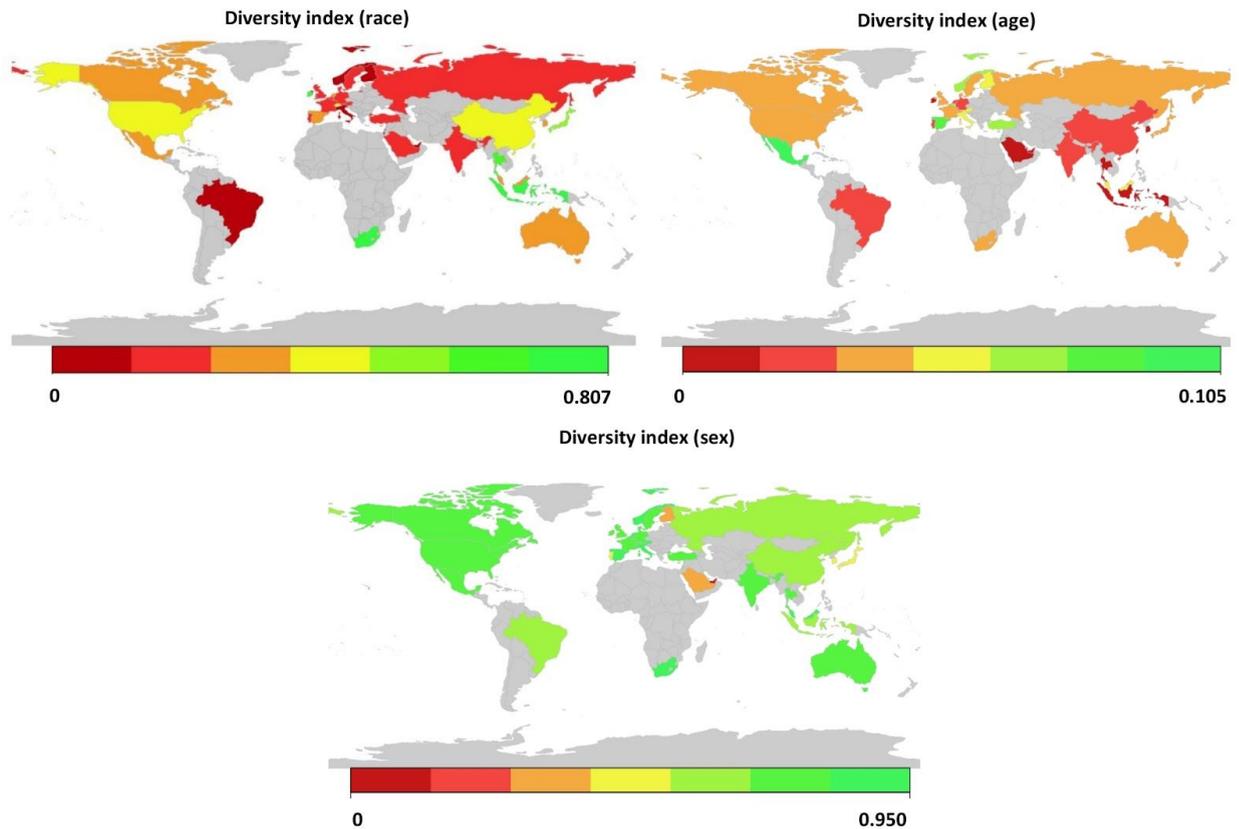

Figure 4. Distribution of median diversity sex, race and age indexes by countries.

**Discussion**

The conflict surrounding increasing diversity is, at least in part, symptomatic of the rapid globalization of the world economy and culture. Despite anti-discrimination efforts undertaken by the UN and other international organizations, many people still experience or witness discrimination in various aspects of their lives. While more and more research provide support towards the driving force of diversity in innovation and business success (Pager, 2008; Hewlett et al., 2013), some companies still lack social variety. Present study was focused on the three most accessible for our method of analysis (image analysis) forms of diversity in business: diversity in the workspace in terms of sex, race and age .

Sexism, discrimination based on sex expression or biological sex, is widely believed to disproportionately be directed towards and negatively affect women (Jary and Jary, 1991). Over the last century, the increasing globalization of the world economy has led to the creation of programs which aim to advocate for and support women's rights in many countries around the world (Ferree and Tripp, 2006). In most countries, women constitute a significant and essential portion of the population according to The World Bank. As shown in this paper, companies with a higher percentage of women among their corporate executives tend to be located in Europe and the United States, whereas companies without female corporate executives are overwhelmingly based in Asia and Saudi Arabia (Fig. 4). This phenomenon is likely the result from a combination of factors: 1) women's rights movements, programs and laws have been widely effective in Europe and the United States, 2) on average, sex- or sex-based bias tends to be less common in European countries and the United States, and 3) there is little to no historical precedent for severe sexism in Europe and the United States as compared to other parts of the world. On the contrary, the latter has explanatory power regarding the preponderance of Asian and Saudi Arabian companies that are "female-free". Unfortunately, given that the percentage of women among corporate executives in the majority of the sampled companies is significantly lower than the percentage of women in the population of the country in which the company is located, it is clear that sex discrimination is more or less inherent among corporate leadership in large multi-national companies. Congruent with this theory, recent evidence suggests that in published text, female names are associated with "career" words significantly less often than male names; demonstrative of the widespread nature of implicit and explicit sex biases (Caliskan et al., 2017).

Racism, characterized by discrimination based on race or ethnicity, is arguably the most common and widespread form of discrimination. One of the most ubiquitous and archetypical forms of racism is directed against people considered to be black (Garner, 2017; Van Dijk, 2015).

The companies with the largest proportion of black corporate leadership were those whose headquarters were in the Republic of South Africa (Table 4, Fig. 4), ostensibly because black citizens comprise a substantial majority of the country's population. However, it is worth noting that even in these companies there is a disparity between the demographics of their corporate leadership and the demographics of the nation. The United States, now host to the largest number of companies in the TOP500 Forbes list (Table 1), still operates under the shadow of a problematic and troubled history of racism (Garner, 2017; Van Dijk, 2015). Nonetheless, black Americans constitute a significant minority (approx. 13 %) of the population of the United States, which is among the largest of any non-African country. Many sampled companies in the United States reflected this demographic statistic in their relatively sizeable proportion of black corporate executives (Table 4). Alternatively, these companies' success could possibly be attributed to the passage of effective civil rights legislation and enforcement of equal protection in the 20[th] Century (Bonilla-Silva, 2001; Garner, 2017; Van Dijk, 2015). The data concerning the proportion of black corporate executives in American companies can be characterized as follows: 1) Despite highly heterogenous ethnic demographics in different American states, there was no correlation between the proportion of black corporate executives and the black population as a percentage of the total state population; in most companies, it was lower than 20 % (Fig. 3D). 2) More than 30 companies have corporate leadership that is devoid of black executives, standing in sharp contrast to the significant attention paid by Americans to the absence of diversity. Despite measures, such as affirmative action, taken by the United States government to promote the equal representation of minorities in the workplace, there continues to be racial stratification among corporate leadership – this may be indicative of immense implicit or explicit bias. The European dataset is characterized by a relatively low proportion of black executives, which is in accordance with the comparatively low proportion of the black population in most European countries. This concurrence among the data renders it difficult to affirm the presence of racially motivated bias among the sampled European companies. The predominant commonality among the sampled Asian companies manifested as limited racial diversity in company leadership – some of these companies' corporate executives were found to be fully ethnically homogeneous. It is important to note, however, that Asian companies represented a significant majority of the companies for which there were no publicly available photographs of corporate executives. The impact of this potentially confounding factor is somewhat subdued by the relative magnitude of the dataset for Asian companies, allowing us to reasonably conclude that the sampled Asian companies are exceedingly marked by ethnic and racial homogeneity – increasing economic globalization notwithstanding.

Representing a divergence from the data regarding sex and race, analysis of the CNN-based age predictions did not reveal any distinct trends among the sampled companies. The median age of the corporate leaders fell between 50 and 52 years old for the vast majority of the sampled companies. The distribution of predicted ages exhibited substantial deviation for the $1^{st}$ quartile values across different companies, which is indicative of the previously described under-estimation of subject age. It is possible that the consistent underestimation is an artifact of variations in the resolution and currency of the photographs. Alternatively, since the pre-trained dataset contained relatively few photographs of Asian people, the algorithm may have been less accurate when predicting the age of Asian subjects. Despite the lack of algorithmically predicted age data, we will attempt to ascertain the severity of age-based bias in corporate leadership.

In summary, thorough analysis of the data as a whole and by geographical region has demonstrated that the collection of corporate leader's photos incorporated implicit sex- and race-based biases. Presumably, any bias present in a training dataset will be imparted upon an algorithm that trains with said dataset. As discussed above, the dataset used for this work over-represents white and male subjects, while underrepresenting non-white and female subjects. It is possible, however, that AI based approaches may be developed to estimate the severity of the underlying bias.

Particularly, the underlying societal issue was addressed by Chisa Enomoto, the Head of the Social Communication Office for the TEPCO Group, who said in regard to the company's actions in furtherance of workplace diversity: "*Like many other Japanese companies, was working to make it easier for women to excel professionally. However, it seems to me that ... efforts to change attitudes among men were severely lacking ... Setting numerical targets to quantitatively develop a new culture is an important form of positive action TEPCO could take to achieve global-level parity on this front*". This sentiment illustrates the concern held by corporate leadership in large companies and their desire to reduce the amount of discrimination in business. At present, the rapid expansion of computing power offers an opportunity to use computer algorithms or AI to analyze and estimate the prevalence and severity of discrimination.

The margin of error present in the predicted data for sex, age and race was in some cases larger than 10 %. This large degree of uncertainty may be attributed to a small sample size for most companies and inconsistent photo quality. In addition, the non-availability of public photographs for every sampled company may have in some way contributed to the error. In this sense, it is still reasonable to expect further development of more accurate and applicable algorithms in the future. Moreover, investigation into other methods or techniques, such as predicting ethnicity and sex via the subject's name (Ryan et al., 2012), appears to hold promise.

The further development of technology in this field depends heavily on the involvement of the company's leadership and their desire to promote their development. It is important to understand whether existing methods of analysis yield an adequate numerical estimation of discrimination, how accurate the predictions of existing algorithms are, if it is necessary to improve them, and whether it is important to provide available qualitative information for quantification of the diversity. Future research may look at correlation between diversity index and company ranking.

**Limitations of the Study**

The authors make no claims regarding the novelty of the methods used in this study. While the authors developed a range of DNN-based image recognition tools, the deep neural networks used for the purposes of this study were developed, trained and published by the other groups referenced in this paper to avoid shifting the focus of this study from the central theme of developing the automated organizational diversity metrics. The dataset has many limitations. Many of the companies from the Forbes Global 2016 2000 list did not have any executive profiles on their websites and in many cases the pictures of the executives were acquired and cross-referenced using Google Images (images.google.com) search engine and verified manually. This may have resulted in a number of errors and mismatches in the dataset. Many of the photographs used in study were edited and various visual effects and filters made it difficult for a human eye and AI algorithms to identify the age, sex and race of executives and may have negatively contributed to classification accuracy.

Another limitation with the dataset and the results of the study is the fluidity of the workforce. For example, the latest version of the dataset was assembled on March 20th, 2017 and Intel Corporation was highlighted as one of the companies with zero African black. On March 29th, 2017 Intel announced the appointment of Aicha S. Evans for a position of Chief Strategy Officer, substantially improving the diversity index.

There is also a limitation of the diversity index, which compares organizational diversity to the population diversity in the company's country of origin. Many of these companies are global multinational companies servicing international markets with substantial workforce located in other countries. A comprehensive diversity index may be developed to evaluate how representative the company's management is of its target audience.

The authors acknowledge that there may be substantial errors in the study since for most companies the outputs were not cross-checked by the human operators. This study does not imply that organizational diversity at the management level is a source of competitive advantage of success of the company, which may be a subject for further investigation.

**Future Directions**

The accuracy of the age, race and sex classifiers may be improved and more classes for the race should be added and the dataset collection and executive profile management methods should be improved. When the complete and accurate datasets of management profiles are available, DNNs may serve as a powerful tool for rapid analysis of diversity on the government level (e.g.

presidential administrations), educational institutions and in the media outlets.

Using AI to automatically assess organizational diversity is a highly controversial topic and one of the young female collaborators, who performed a substantial amount of work asked to be removed from the study after being intimidated by a journalist from a media outlet with a low diversity index. The study may benefit from a crowd sourced and crowd curated effort to improve the quality of the datasets and automate the process of assessing the organizational diversity and developing the reporting and recommendation systems for the relevant stakeholders.

**Acknowledgements**

We would like to thank our editor Kainen L. Utt for his advice and suggestions regarding the clarity of this work. We would like to acknowledge the dedicated technical assistance of Mr. Anuar Taskynov.